\documentclass[aps,prl,reprint,showpacs,superscriptaddress,nobibnotes,nofootinbib]{revtex4-2}
\usepackage{graphicx}
\usepackage{amssymb} 
\usepackage{amsmath}
\usepackage [colorlinks=true, linkcolor=blue, citecolor= blue, urlcolor=blue,]{hyperref}
\begin{document}

\title{Reanalysis of the $\beta - \bar{\nu_e}$ Angular Correlation Measurement from the aSPECT Experiment with New Constraints on Fierz Interference}
%%%%%%%%%%%Author list
\author{M. Beck}
\affiliation{Institut f\"{u}r Physik, Johannes Gutenberg-Universit\"{a}t, 
	55128 Mainz, Germany}
\author{W. Heil}
\email{Corresponding author: wheil@uni-mainz.de}
\affiliation{Institut f\"{u}r Physik, Johannes Gutenberg-Universit\"{a}t, 
	55128 Mainz, Germany}
\author{Ch. Schmidt}
\affiliation{Institut f\"{u}r Physik, Johannes Gutenberg-Universit\"{a}t, 
	55128 Mainz, Germany}
\author{S. Bae\ss ler}
\affiliation{Department of Physics, University of Virginia, Charlottesville, Virginia 22904, USA and Oak Ridge National Lab, Bethel Valley Road, Oak Ridge, Tennessee 37831, USA}

\author{F. Gl\"uck}
\affiliation{Institut für Astroteilchenphysik (IAP), Karlsruhe Institute of Technology (KIT), 76344 Eggenstein-Leopoldshafen, Germany}

\author{G. Konrad}
\affiliation{Technische Universit\"at Wien, Atominstitut, 1020 Wien, Austria}

\author{U. Schmidt}
\affiliation{Physikalisches Institut, Ruprecht-Karls-Universit\"{a}t, 69120 
Heidelberg, Germany}

\date{\today}
%%%%%%%%%%%%%Abstract
\begin{abstract}
On the basis of revisions of some of the systematic errors, we reanalyzed the electron-antineutrino angular correlation ($a$ coefficient) in free neutron decay  inferred from the recoil energy spectrum of the protons which are detected in 4$\pi$ by the aSPECT spectrometer. With $a = -0.10402(82)$ the new value differs only marginally from the one published in 2020. The experiment also has sensitivity to $b$, the Fierz interference term. From a correlated $(b,a)$ fit to the proton recoil spectrum, we derive a limit of $b = -0.0098(193)$ which translates into a somewhat improved 90\% CL region of $-0.041 \leq b \leq 0.022$ on this hypothetical term. Tighter constraints on $b$ can be set from a \mbox{combined$^{(c)}$} analysis of the PERKEO~III ($\beta$ asymmetry) and aSPECT measurement which suggests a finite value of $b$ with $b^{(c)}$ = -0.0181 $\pm$ 0.0065 deviating by 2.82$\sigma$ from the standard model. 
\end{abstract}

%\pacs{06.30.Ft, 07.55.Ge, 11.30.Cp, 11.30.Er, 04.80.Cc, 32.30.Dx, 82.56.Na}

\maketitle

%%%%%%%%%%%%%main
The free neutron presents a unique system to investigate
the standard model (SM) of particle physics. While the neutron lifetime gives the overall strength of the
weak semileptonic decay, neutron decay correlation coefficients depend on the ratio of the coupling constants involved, and hence determine its internal structure.
The aSPECT experiment \cite{a1,a2,a3} has the goal to determine the ratio of the weak axial-vector and vector coupling constants $\lambda =g_\text{A}/g_\text{V}$ from a measurement of the $\beta - \bar{\nu_e}$ angular correlation in neutron decay. The $\beta$-decay rate when observing only the electron and neutrino momenta and the neutron spin and neglecting a $T$-violating term is given by \cite{a4}
\begin{align} \label{equ:1} 
\nonumber
d^3\Gamma=&G_\text{F}^2V_\text{ud}^2\left(1+3\lambda^2\right)p_eE_e\left(E_0-E_e\right)^2\\\nonumber
&\cdot \left(1+a\frac{\vec{p_e}\cdot\vec{p_\nu}}{E_eE_\nu}+b\frac{m_e}{E_e}+\frac{\vec{\sigma_n}}{\sigma_n}\cdot \left[A\frac{\vec{p_e}}{E_e}+B\frac{\vec{p_\nu}}{E_\nu}\right] \right)\\
&\cdot dE_ed\Omega_e d\Omega_\nu
\end{align}
with $\vec{p_e}$, $\vec{p_\nu}$, $E_e$, $E_\nu$ being the momenta and total energies of the beta electron and the electron-antineutrino, $m_e$ the mass of the electron, $G_\text{F}$ the Fermi constant, $V_\text{ud}$ the first element of the Cabbibo-Kobayashi-Maskawa (CKM) matrix, $E_0$ the total energy available in the transition, and $\vec{\sigma_n}$ the spin of the neutron. The quantity $b$ is the Fierz interference coefficient. It vanishes in the purely vector axial-vector ($V-A$) interaction of the SM since it requires left-handed scalar ($S$) and tensor ($T$) interactions (see below). The correlation coefficients $a$ and $A$ ($\beta$-asymmetry parameter \cite {a5,a6}) are most sensitive to $\lambda$ and are used for its determination. The SM dependence of the electron-antineutrino angular correlation coefficient $a$ on $\lambda$ is given by \cite {a4,a7,a8} 
\begin{equation} \label{equ:2} 
a=\frac{1-|\lambda|^2}{1+3|\lambda|^2}.
\end{equation}
In short, at aSPECT the $a$ coefficient is inferred from the energy spectrum of the recoiling protons from the $\beta$ decay of free neutrons. The shape of this spectrum is sensitive to $a$ and it is  measured in $4\pi$ by the \mbox{aSPECT} spectrometer using magnetic adiabatic collimation with an electrostatic filter (MAC-E filter) \cite {a9,a10}. This technique in general offers a high luminosity combined with a well-defined energy resolution at the same time. In order to extract a reliable value of $a$, any effect that changes the shape of the proton energy spectrum, or to be more specific - the integral of the product of the recoil energy spectrum and the spectrometer transmission function
- has to be understood and quantified precisely. With the analysis of all known sources of systematic errors at that time and their inclusion in the final result by means of a global fit, the aSPECT Collaboration published the value $a= -0.104 30(84)$  \cite {a11}. From this, the ratio of axial-vector to vector coupling constants was derived giving $|\lambda|  = 1.2677(28)$. 
This value deviates by $\approx 3 \sigma$ from the most precise PERKEO III result \cite{a5}, determined via the $\beta$-asymmetry parameter $A$. One possible explanation for this discrepancy are hypothetical scalar or tensor couplings in
addition to the $V-A$ interaction of the SM. In this case, the overall decay probability may be modified by the Fierz interference term $b$ according to Eq.~(1) which would also change the measured values of many of the correlation coefficients; see also Refs. \cite{f1,f2}.   On the other hand, SM differences might be of experimental origin, which generally requires a critical reexamination of the systematics in the respective datasets. Consequently, the  aSPECT data were reanalyzed as it could not be ruled out that  previously overlooked systematics in i) backscattering and below threshold losses in the detector and ii) the retardation voltage $U_\text{AP}$ of the electrostatic filter  could be the reason for the discrepancies in the $\lambda$ values. With these revised systematics included in the global data fit, a new SM analysis of the correlation parameter $a$ is performed as well as a combined $(a,b)$ analysis in order to put a direct constraint on the Fierz interference term from a single measurement. Tighter constraints on $b$ can be set by using the PERKEO~III data, where limits on $b$ have been derived via a combined $(b,A)$ fit to their data.

{\bf Backscattering and below threshold losses:} 
Whereas the amount of electron-hole pair production in amorphous solids by a penetrating proton can be determined rather accurately with the binary collision code TRIM \cite {a12}  (used in \cite {a11}), the calculation of the ionization depth profile in crystalline solids is more complicated due to channeling effects that TRIM does not attempt to take into account. In our reanalysis we simulated the slowing down of protons in our silicon drift detector (SDD) (processed on a $\left<100\right>$-oriented Si wafer) by the program Crystal-TRIM originally developed in order to describe ion implantation into crystalline solids with several amorphous overlayers \cite{a13,a14} (in our case: a 30 nm thick aluminum overlayer including its 4~nm thick alumina layer \cite{b1} on top). The range of applicability of this code was studied by comparing with existing molecular dynamics simulations.\footnote {Molecular dynamics (MD) methods are well suited to study ion penetration in materials at energies where also multiple simultaneous collisions may be significant. The MDRANGE code \cite{MDRANGE}, however, requires too extensive computation time given the high number of particle tracking simulations ($\approx 2\cdot 10^7$ protons).} With Crystal-TRIM good agreement (5\%) was obtained for the parameter $C_\text{el}=0.65$ in the semiempirical formula for the local electronic energy loss \cite {a14}.

Together with the charge collection efficiency  for this type of detector \cite{a18} at depth $z$ and further taking into account charge exchange reactions of backscattered protons at the topmost detector layer (not considered so far) \cite{Phillips}, we derive our simulated pulse height spectra. The latter ones are in excellent agreement with the experimental pulse height spectra at different acceleration voltages $U_\text{acc}$ and retardation voltages $U_\text{AP}$. This procedure allows us to calculate the below-threshold losses including the events with no energy deposition inside the detector. More details are presented in the Appendix.\\
Figure~1 shows the fractional losses for the two detector pads. In both cases a cubic spline interpolation was used to describe their retardation voltage dependence. In total, channeling and the inclusion of charge exchange reactions do not significantly influence the spectral shape of the undetected protons, as seen by comparison with Fig. 22 in \cite{a11}. The obvious higher fractional losses of $\approx 35\%$ (shape independent), which can be attributed to the equilibrium charge ratio of backscattered protons from the alumina layer, have no effect on the final result due to normalization ($N_\text{0}$).\\
  
\begin{figure}
    \centering
    \includegraphics[width=0.48\textwidth]{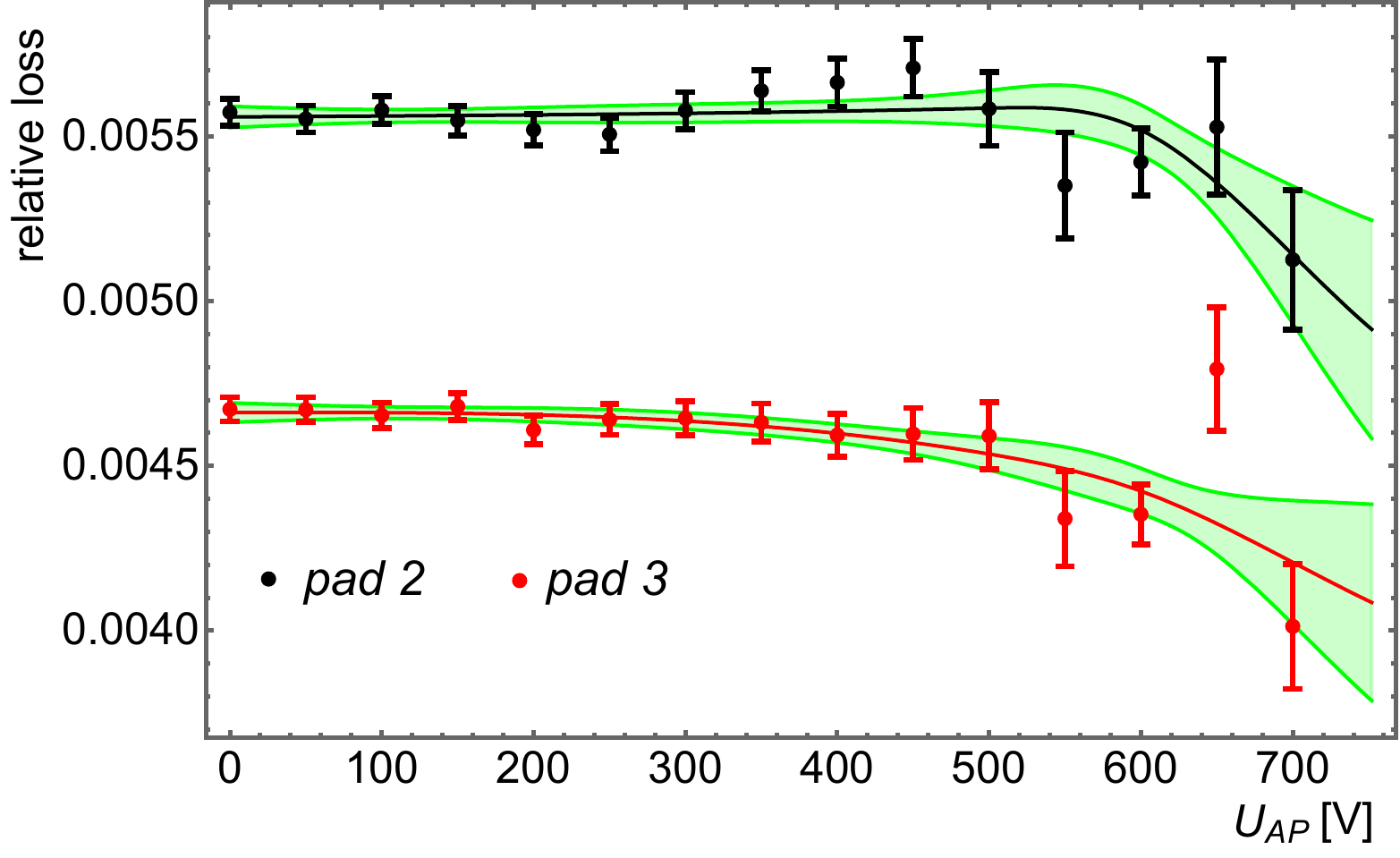}
    \caption{ Simulated fraction of undetected protons (both for pad 2 and pad 3 of the SDD) of the integral proton spectrum whose corresponding pulse heights fall below the threshold of the DAQ system. The $U_\text{AP}$ dependence of the simulated losses can be described by a cubic spline function (solid line) and is shown together with the $1\sigma$ error predicted by the global fit (green band). 
    }    \label{fig:1}
\end{figure}
{\bf Effective retardation voltage $\left<U_\text{A}\right>$:}\\
Like the magnetic field ratio $r_\text{B}=B_\text{A}/B_0$, where $B_0$ and $B_\text{A}$ are the respective magnetic fields at the place of emission and retardation, the retardation voltage $U_\text{AP}$ directly enters the spectrometer’s adiabatic transmission function \cite{a2}\footnote{The transmission function determined  from particle tracking simulations \cite{a11} (with the only input variable: the precisely known electromagnetic field of aSPECT) agrees with the analytical description of the spectrometer properties, i.e., protons move adiabatically through the MAC-E filter.} given by
\begin{equation}\label{equ:3}
\begin{split}
F_{\text{tr}}&(T,U_\text{AP},r_\text{B}) \\
& = \left \{ \begin{array}{ll}
{0}&\text{if } T\leq eU_\text{AP}\\
{1-\sqrt{1-\frac{T-eU_\text{AP}}{r_\text{B}T}}}&\text{if }eU_\text{AP}<T<\frac{eU_\text{AP}}{1-r_\text{B}}\\
{1}&\text{if } T>\frac{eU_\text{AP}}{1-r_\text{B}}\\
\end{array}\right.
\end{split} ,
\end{equation}

where $T$ is the kinetic energy of the isotropically emitted protons.
The inhomogeneities of the potential in the decay volume (DV) and the analyzing plane (AP) region result in a slight shift (determined from
particle tracking simulation) of the effective retardation voltage $\left<U_\text{A}\right>$ from the applied voltage $U_\text{AP}$ due to spatial and temporal variations of the work function of the DV and AP electrodes. This shift and its functional dependence on $U_\text{AP}$ is described in detail in \cite{a11}.
A further source of uncertainty is the measurement precision $\sigma_\text{AP}^\text{Agilent}$ of the applied voltage by means of the Agilent 3458A multimeter. In the previous analysis this uncertainty was not correctly incorporated in the fit function as an error of the horizontal axis in the $U_\text{AP}$ dependence of the integral proton spectrum. Instead, it was treated as part of an offset error $c_\text{offset}^{\left<U_\text{A}\right>}$ common to all $\left<U_\text{A}\right>$ values. We now improve our analysis by replacing Eq.~(3) through
\begin{equation}\label{equ:4} 
F_{\text{tr}}(T,U_\text{AP},r_\text{B}) \longrightarrow  
F_{\text{tr}}(T,\left<U_\text{A}\right>,r_\text{B})+ \frac{\partial F_{\text{tr}}}{\partial U_\text{AP}} \Delta U_\text{AP}^\text{Agilent}~.
\end{equation}
Here, $\sigma_\text{AP}^\text{Agilent}$ enters the transmission function via the partial derivative $\Delta F_{\text{tr}}^\text{Agilent}=(\partial F_{\text{tr}}/\partial U_\text{AP}) \Delta U_\text{AP}^\text{Agilent}$~.
In the fit procedure, $\Delta U_\text{AP}^\text{Agilent}$ is a restricted fit parameter, Gaussian distributed with zero mean and $\sigma_\text{AP}^\text{Agilent}= 13~\text{mV}$.

\section{Global fit results}
In the ideal case without any systematic effect, the fit to the proton integral count rate spectrum would be a $\chi^2$ minimization of the fit function
\begin{equation}\label{equ:5} 
f_\text{fit}(U_\text{AP},r_\text{B};a,N_0)=N_0\int_0^{T_\text{max}}\omega_\text{p}(T,a)F_{\text{tr}}(T,U_\text{AP},r_\text{B})dT
\end{equation}
with the overall prefactor $N_0$ and $a$ as free fit parameters. The way to include all systematic corrections to the global fit, the reader is advised to refer to the relevant Sec. III C of \cite{a11} for details. The theoretical proton recoil spectrum $\omega_\text{p}(T,a)$ is given by Eqs.~(3.11,3.12) in \cite{a24} with $\lambda$ replaced by $a$ [Eq.~(2)]. This spectrum includes relativistic recoil and higher order Coulomb corrections, as well as order-$\alpha$ radiative corrections \cite{a24,a25}\footnote{In contrast to \cite{a24}, the relative radiative correction $r_\text{p}$ \cite{a25} to the proton energy spectrum contains additional decimal places that have no effect on the measurement accuracy of $a$ achieved \cite{a11}.}. For the precise computation of the $F(Z = 1,E_e)$ Fermi function (Coulomb corrections), we have used formula (ii) in Appendix 7 of Ref.~\cite{a26}. The weak magnetism $\kappa=(\mu_p-\mu_n)/2$ (conserved vector current value) is included in the $(E_p, E_e)$ Dalitz distribution, but essentially drops out in the proton-energy spectrum after integration over the electron energy $E_e$ \cite{a27}.  All corrections taken together are precise to a level of $\Delta a/a\approx0.1\%$.\\
The aSPECT experiment also has sensitivity to $b$, the Fierz interference term [see Eq.~(1)]. The $ 1/E_e$ dependency results in small deviations from the SM proton recoil spectrum. In Eqs.~(4.10, 4.11) of \cite{a28} an analytical expression of $\omega_\text{p}^*(T,a)$ is given where recoil-order effects and radiative corrections are neglected$^{(*)}$\footnote{In complete agreement with Eqs.(9-15) of Ref. \cite{f1}.}. To add the $b$ term to the complete spectrum $\omega_\text{p}(T,a)$, we define $\omega_\text{p}(T,a,b)$ by adding an additional term  $4/(1+3a)m_e E_e(E_{\text{2m}}-E_e/2)\cdot b$ to the right-hand side  of Eq.~(3.12) in \cite{a24}, and replace $\Delta=m_n-m_p$ by $E_{\text{2m}}=\Delta-(\Delta^2-m_e^2)/(2m_n)$. For the final result, we performed a global fit as described in Sec. V of \cite{a11}. Table~I summarizes the  results on $a$ from the purely SM approach, i.e., $\omega_\text{p}(T,a)$ as well as the simultaneous ($b,a$) fit results if the proton recoil spectrum $\omega_\text{p}(T,a,b)$ would be modified by the Fierz interference term $b$.\\
The error on $a$ from the respective fit is the total error scaled with $\sqrt{\chi^2/\nu}$. Besides the statistical error, it contains the uncertainties of the systematic corrections and the correlations among the fit parameters which enter the variance-covariance matrix to calculate the error on the derived quantity from the fit. In \cite{a11} we stated that the elevated $\chi^2/\nu$ values of the global-$a$ fit most likely arise due to the nonwhite reactor power noise and/or high-voltage induced background fluctuations (cf. Sec. IV A there). The reanalysis of the aSPECT data  now leads to a reduced  $\chi^2/\nu=1.25$ ($p =4.1\cdot10^{-3}$ for $\nu =264$). The revision of the $U_\text{AP}$ error is the main driver for this and for the corresponding changes to $a$.\\
Our new value for the $a$ coefficient only differs marginally from the one published in \cite{a11} (see Table I) and is given by
\begin{equation}\label{equ:6} 
a=-0.10402\pm0.00082~~.
\end{equation}
Using Eq.~(2) we derive for $\lambda$ the value  $\lambda = -1.2668(27)$.\\

If one allows for $b$ as free parameter, we obtain from the measurement of the proton recoil spectrum a limit at 68.27\% CL for the Fierz interference term of
\begin{equation}\label{equ:7} 
b=-0.0098\pm0.0193~.
\end{equation}
In the combined fit, the error on $a$ increases by a factor of 1.7 as compared to the SM
analysis with $b\equiv0$ (Table 1), since the two fit parameters show a fairly strong correlation: the off-diagonal element $\rho_{a,b}$ of the correlation matrix is $\rho_{a,b}=0.808$ as a result of the global fit. Our limit can be rewritten as $-0.041\leq b\leq 0.022$ ($90\%$ CL) which is currently the most precise one from neutron $\beta$ decay \cite{a31}.
\begin{table*}
\centering
\resizebox{0.75\textwidth}{!}{%
	\begin{tabular}{lcccccc} 
		\hline \hline
		%dir & $f$/Hz & $B_\text{ex}$/µT &$A_\text{x}$/nT&$\phi_\text{x}$  &$A_\text{y}$/nT&$\phi_\text{y}$ &$A_\text{z}$/nT&$\phi_\text{z}$ \\[.5ex]
		&$a$&$\Delta a$&$b$&$\Delta b$&$\chi^2/\nu$&$p$-value\\
		\hline
		Results from \cite{a11}&$-0.10430$&$0.00084$&-&-&$1.440 (\nu=268)$&$3.1\cdot10^{-6}$\\
	
		Reanalysis&$-0.10402$&$0.00082$&-&-&$1.245 (\nu=264)$&$4.1\cdot10^{-3}$\\
		
		$(a,b)$ analysis&$-0.10459$&$0.00139$&$-0.0098$&$0.0193$&$1.249 (\nu=263)$&$3.7\cdot10^{-3}$\\
		\hline \hline
\end{tabular}}
\caption{Global fit results (at 68.27\% CL) on $a$ assuming the SM vector and axial-vector couplings only, together with an analysis allowing nonzero scalar or tensor interactions in order to derive limits on $b$. The error bars on $a$ and $b$  were scaled with $\sqrt{\chi^2/\nu}$ whenever the condition $p=\int_{\chi^2/\nu}^\infty f_\nu(\chi^2)d\chi^2<0.05$ was met with $f_\nu(\chi^2)$ being the $\chi^2$-distribution function with $\nu$ degrees of freedom \cite{a30}.}
\label{tab:1}
\end{table*}
%%%%%%%%%%%%%%%%%%%%%%%%%%%%%%%%
\section{Combined analysis of recent measurements of PERKEO III and aSPECT}
Further constraints on $b$ can be set by using the PERKEO~III data, where comparable limits on $b$ have been derived from the measurement of the $\beta$ asymmetry in neutron decay via a combined $(b,A)$ fit to their data (see Fig.~3 in \cite{a32}). Including other measurements like UCNA \cite{hick,a34}, aCORN \cite{a33,b3}, and PERKEO~II \cite{a35} would not add substantial information. Besides, error ellipse data (see Fig.~2) are not (or not yet) published. To allow a direct comparison, the neutron decay parameters $A$ and $a$ are expressed in terms of $\lambda$ (see e.g. \cite{a8}). Figure 2 shows the error ellipses in the ($b,\lambda$) plane which represent the iso-contours of the respective bivariate Gaussian probability distributions (PDF) to visualize a 2D confidence interval \cite{a36}. Both experiments on their own show a value of $b$ compatible with zero in the $1\sigma$ range of their error ellipses ($p_\text{cl}=39.35\%$). With $\rho_{A,b}= -0.985$, the PERKEO~III ellipses have a very strong negative correlation and are almost orthogonal to the ones from the aSPECT $(b,a)$ analysis.  This orthogonality in turn leads to a stronger constraint in $b$ as can be seen directly from the overlap of error ellipses their assigned confidence levels ($p_{\text{cl}}$) are beyond $p_{\text{cl}} > 40\%$. From the combined error ellipse, which represents the iso-contour of the product of the respective PDFs, we can deduce that the case $b=0$ lies on the edge of its 98.1\% confidence region. The resulting values for ($b,\lambda$) at 68\% CL in combining$^{(c)}$ the independent datasets of PERKEO~III and aSPECT are
\begin{equation}\label{equ:8} 
\begin{split}
b^\text{(c)}=-0.0181\pm0.0065~~~\\
\lambda^\text{(c)}=-1.2724\pm0.0013~.
\end{split}
\end{equation}
With $2.82\sigma$, the Fierz interference term $b^\text{(c)}$ obtained deviates from zero while  $\lambda^\text{(c)}$  lies in between the derived PERKEO~III and aSPECT values for $\lambda$ from the prior SM analysis \cite{a5,a11}. Note that the most accurate results for  $A$ and $a$ differ in their derived $\lambda$ values by 3.4 standard deviations within the SM approach (see Fig.2).\\
In order to check that the PERKEO~III ($P$) and aSPECT ($A$) measurements of ($b,\lambda$) are statistically compatible with the combined result given in Eq.~(8), we used the generalized least squares (GLS) method \cite{a37} for fitting.  Taking the 4$\times$4 covariance matrix 
\begin{equation}
\Omega=\begin{pmatrix} \Sigma(P) & 0\\ 0 & \Sigma (A)\end{pmatrix}
\end{equation}
and the 4-vector $V^\text{T}=(b^{(P)}-b^\text{(c)},\lambda^{(P)}-\lambda^\text{(c)},b^{(A)}-b^\text{(c)},\lambda^{(A)}-\lambda^\text{(c)})$, $\chi^2=V^\text{T}\Omega^{-1}V$ was minimized with $b^\text{(c)}$ and $\lambda^\text{(c)}$ as free parameters ($\nu=2$). As input, we took the known quantities $b^{(P)}$, $\lambda^{(P)}$, $\Sigma(P)$ and $b^{(A)}$, $\lambda^{(A)}$, $\Sigma (A)$ from the respective measurements with $\Sigma$ denoting the 2$\times$2 variance-covariance matrix. With $\chi^2_{\nu=2}=3.6$ (goodness of fit test), we arrive at the same results as in Eq.~(8). The resulting $p$-value is $p=0.16$ and is above the threshold of significance (typically 0.05 \cite{a30}). In Fig. 2, the error ellipses with the respective confidence level $p_{\text{cl}}(A)=0.31$ and $p_{\text{cl}}(P)=0.76$ are drawn which both touch in the center of the combined error ellipse. The $p$-value of 0.16 is reproduced by taking the product $(1-p_{\text{cl}}(A))\cdot (1-p_{\text{cl}}(P))$ \cite{a38}.
\begin{figure}[h!]
    \centering
    \includegraphics[width=0.48\textwidth]{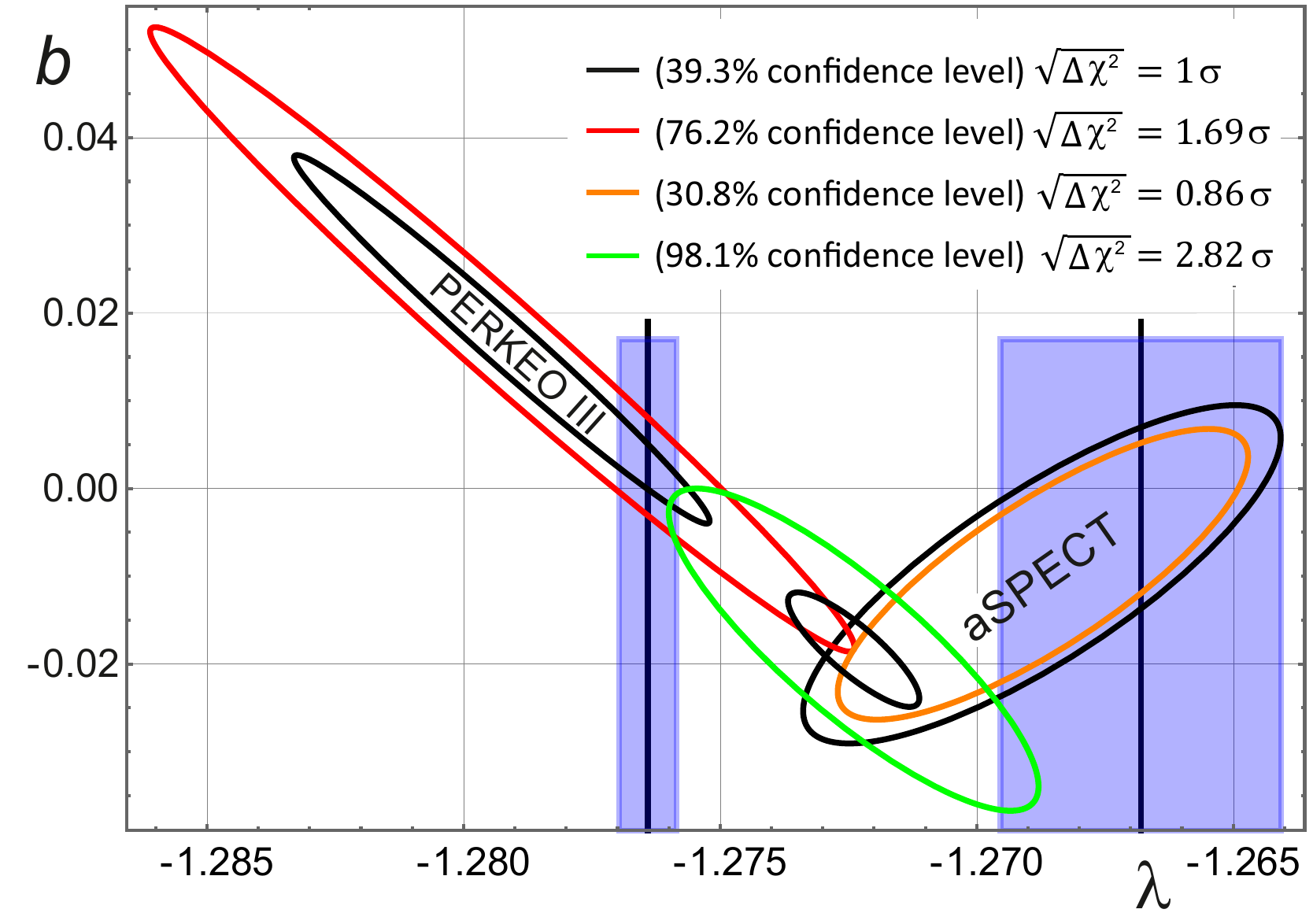}
    \caption{Confidence region for the ratio $\lambda=g_\text{A}/g_\text{V}$ and the Fierz interference parameter $b$. The results of a correlated $(b,\lambda)$ analysis of the PERKEO~III and aSPECT measurements are shown as error ellipses representing specified confidence levels. The combined$^{(c)}$ result from the two independent datasets suggests a finite value of $b$ with $b^\text{(c)}=-0.0181(65)$ which deviates by $2.82\sigma$ from the SM (shown by the green error ellipse). The 90\% CL interval of $-0.012 < b < +0.144 $ from UCNA \cite{a34} is consistent with $b^\text{(c)}$ or a vanishing $b$ value. For the SM analysis ($b\equiv0$) of the two measurements, the $\lambda$ values of PERKEO~III and aSPECT disagree by 3.4 standard deviations (blue vertical bars: $1\sigma$ error).    } 
    \label{fig:1}
\end{figure}

The nonzero value of the Fierz interference term $b^\text{(c)}$ in neutron $\beta$ decay is in tension with constraints from low energy precision $\beta$-decay measurements (pion \cite{a40,a39,cirigliano}, neutron, and nuclei \cite{a41,hardy} ) as well as Large Hadron Collider (LHC) constraints through the reaction $pp\rightarrow e\nu + X$ and $pp\rightarrow e^+e^- + X$ \cite{a42}. The $b^\text{(c)}$ and $\lambda^\text{(c)}$ values of Eq.~(8) predict the neutron lifetime value of $\tau_n^\text{(c)}=(894.2\pm{4.2})~\text{s}$ using the master formula from \cite{a43}\footnote{For $|V_{\text{ud}}|$ we took the PDG value \cite{a31}.} multiplied by the factor $(1+\left<m_e/E_e\right>\cdot b)^{-1}$ \cite{b6} on the right-hand side.
This differs by 3.7 $\sigma$ from the PDG value for the neutron lifetime  $\tau_n=(878.4\pm0.5)~\text{s}$ \cite{a31}.

As shown by Falkowski $et~al.$ \cite{a41}, the SM difference could also be attributed to right-handed couplings for tensor currents
($C_T = -C_T'$ 
and $b_\text{Fierz}=0$).
By taking the PDG values for the neutron lifetime and the decay parameters $A$ and $B$ \cite{a31}, including our result on $a$ [Eq.~(6)], a ($\lambda ,|C_\text{T}/C_\text{A} |$) fit to the data expressed in terms of the Lee-Yang Wilson coefficients \cite{a41} shows a striking preference for a nonzero value of the beyond-SM parameter $|C_\text{T}/C_\text{A} |$ with $|C_\text{T}/C_\text{A} |= 0.047\pm 0.018$ . While this result lies within the recent low energy limits  $|C_\text{T}/C_\text{A} | < 0.087$ (95.5\%~CL) of \cite{b4}, the LHC bounds $|C_\text{T}/C_\text{A} | =4 g_T  | \tilde{\epsilon_T}/{g_A} |< 1.3\cdot10^{-3}$ from  $pp\rightarrow e\nu + X$ \cite{a42,a44} are more stringent than those from ${\beta}$ decays.              

%%%%%%%%%%%%%%%%%%%%%%%%%%%%%%%%
\section{Conclusion and Outlook}
In this paper, we presented a reanalysis of the \mbox{aSPECT} data with an improved tracking of some of the systematic errors.  The value $a = -0.10402(82)$ differs only marginally from the one published in \cite{a11}. The \mbox{aSPECT} experiment has sensitivity to $b$. We extract a limit of $b= -0.0098(193)$ on the Fierz interference term from a combined $(b,a)$ analysis of the proton recoil spectrum. The apparent tension to the PERKEO~III result \cite{a5} based on the SM analysis can be resolved by combining the results of the $(b,\lambda)$ analyses from these two measurements. The finite value for the Fierz interference term of $b^\text{(c)}=-0.0181(65)$ deviates by $2.82\sigma$ from the SM. The goodness of fit test shows that the $(b,\lambda)$ data from PERKEO~III and aSPECT are statistically compatible with the combined result. 
The upcoming Nab experiment \cite{a47} and the next generation instruments like PERC \cite{a48,a49} will allow the measurement of decay correlations with strongly improved statistical uncertainties to underpin these findings or to establish that the SM differences are of experimental origin.
%%%%%%%%%%%%%%%%%%%%%%%%%%%%%%%%%%
\section{Appendix: Backscattering and below threshold losses}
%%%%%%%%%%%%%%%%%%%%%%%%%%%%%%%%%%
Protons reaching the silicon drift detector (SDD) (processed on a $\left<100\right>$-oriented Si wafer \cite{Simson1,Simson2}) can get backscattered due to scattering off the nuclei of the detector material.
Consequently, these protons deposit only a fraction of their kinetic energy inside the active detector volume and the
resulting pulse height may fall below the threshold of the DAQ system. Backscattering depends on the energy of the incident proton, $15 ~\text{keV} < (E_\text{kin} = 15 ~\text{keV} + T) < 15.75 ~\text{keV}$, and its impact angle. The distribution of both quantities is affected by the applied retardation voltage, $U_\text{AP}$· The $U_\text{AP}$ dependence of the detection efficiency may change the value of the $\beta - \bar{\nu_e}$ angular correlation coefficient $a$ extracted from the integral proton spectrum \cite{a11}. \\

The protons relevant for aSPECT have a very short range in the SDD (516~nm: extreme case of channeling), thus the efficiency for proton detection is extremely sensitive to the detector properties near the surface. A proton penetrating the detector first has to pass the amorphous entrance window, which is comprised of a $\Delta z = 30 ~\text{nm}$ thick aluminum overlayer including its $4~\text{nm}$ thick alumina layer \cite{b1} on top. Free charge carriers produced by the proton in this region will not be detected. Even after the entrance window, not all charge carriers will be collected in the central anode of the SDD. The charge-collection efficiency at the border (z' = 0) between the entrance window and active silicon bulk is $\approx$ 50\% and rises with increasing depth ($z' > 0$) according to the following equation \cite{a18}:
\begin{equation}\label{equ:S1}
\renewcommand{\theequation}{S.1}
f_{\text{CCE}}~(z) = \begin{array}{ll}
{0}&\text{for}~~ z' = z-\Delta z < 0\\
S+B{(\frac{z'}{L})}^c &\text{for}~~ 0\leq z'\leq L\\
{1-A~\text{exp}~(-\frac{z'-L}{\tau})}&\text{for}~~ L < z' \leq D
\end{array}\\
\end{equation}
\begin{equation}\nonumber
\text{with}~~A = (1-S)\frac{\tau c}{L+\tau c}~~B = (1-S)(1-\frac{\tau c}{L+\tau c})~.
\end{equation} The total thickness $D$ of the detector is about 450~\text{$\mu$m}, much thicker than the maximum penetration depth of low energy protons.  In order to determine the four remaining parameters, $S, ~c, ~\tau, \text{and} ~L$ of the
charge-collection efficiency function, one has to calculate the effective deposited ionization energy of each proton,
\begin{equation}\label{equ:S2}
\renewcommand{\theequation}{S.2}
E_{\text{dep}}^{\text{C-TRIM}}=\sum_i E_{\text{ion.}}^{\text{C-TRIM}}(z_i)f_{\text{CCE}}(z_i)~~,
\end{equation}
a quantity which is proportional to the measured pulse height. Hence, the histogram of $E_{\text{dep}}^{\text{C-TRIM}}$ from many simulated proton events reproduces our pulse height spectra, if the parameters of $f_\text{CCE}$ are correct. \\

In \cite{a11}, the ionization depth profile was
determined by the TRIM code \cite{a12} a standard method to calculate the interactions of energetic ions with amorphous targets; here in particular the amount of electron-pair production in silicon by a penetrating proton. Due to the crystalline structure of silicon, the underlying assumption for amorphous targets that the impact parameter
distribution is independent of the relative orientation of the target to the beam direction is not valid any more.
Thus, when energetic ions are moving in crystals, they may penetrate much deeper if
they happen to be directed in some specific crystal directions. This ‘channeling’ effect influences the overall trajectory of the impinging particle and causes a strong reduction of the backscattering yield from the lattice atoms \cite{Grahmann}. An issue that may lead to an as yet unidentified systematic in the extraction of $a$.\\

In our reanalysis of the data, the ionization depth profile $E_{\text{ion.}}^{\text{C-TRIM}}(z_\text{i})$ [see Eq.~(S.2)] is determined by the  binary collision code Crystal-TRIM which describes ion implantation into crystalline solids with several amorphous overlayers \cite{a13,a14}.
The respective penetration depths were divided into bins of 3~nm and the deposited
ionization energy per bin was determined including the amorphous overlayer with $f_{\text{CCE}} (z_i<30~\text{nm})=0$.
The range of applicability of this code was studied by comparing with existing molecular dynamics simulations \cite{NordlundPRB} which reproduce the experimental ion penetration depth profiles in materials like Si at ion energies where also multiple simultaneous collisions may be significant.
In order to fix the empirical parameter $C_\text{el}$ in the formula for the local electronic energy loss \cite {a14}, Crystal-TRIM simulations on a Si-cell (diamond structure) with a [100] surface normal were performed, whereby tilting ($\theta$) and twisting ($\varphi$) the direction of the incomimg 10~keV H-ions in the angular range relevant to us ($ 0^{o} \leq \theta \leq 25^{o}~\text{and}~0^{o} \leq \varphi \leq 90^{o} $). We have chosen 10~keV H-ions to allow a direct comparison to existing MDRANGE code depth profiles in Si and to be as close as possible to the true kinetic energy $E_\text{kin}$ of the protons incident on our SDD. Figure S1 shows the mean range of the H-ions depicted in a polar diagram where channeling effects are clearly visible: the impinging beam sees different crystalline planes and axis as the polar and azimuthal angle changes. With Crystal-TRIM, good agreement (5\%) is achieved with the corresponding MDRANGE code profiles (see Fig.~6 in \cite{NordlundPRB}) for the empirical parameter $C_\text{el}=0.65$. About the same value is determined by extrapolation to Z=1 (H-ion) from recommended values of  $C_\text{el}$ for B-, F- and As-ions with 15 keV incident energy \cite{PosseltCTRIM}.\\
\begin{figure}
    \centering
    \renewcommand{\thefigure}{S1}
          \includegraphics[width=0.48\textwidth]{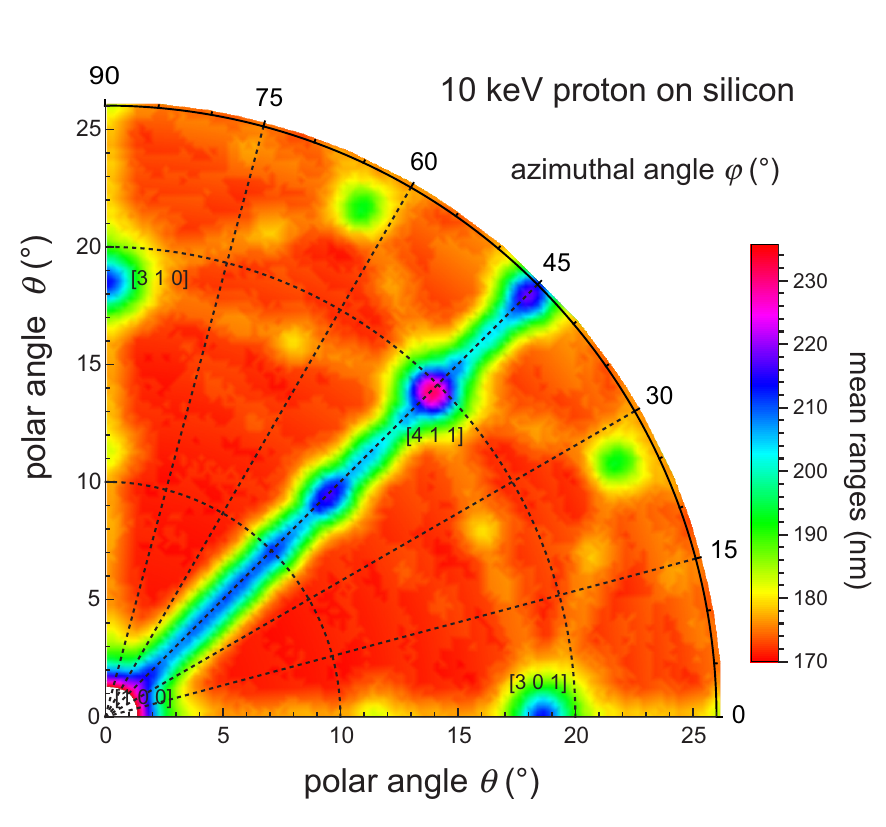}
    \caption{Channeling effects (simulated by Crystal-TRIM) in the range distribution of 10~keV H-ions in Si if they happen to be directed in some specific crystal directions (polar diagram). Miller indices are shown for some of the principal crystal directions. The colors show the mean ranges of the H-ions (white: beyond the displayed color range). For symmetry reasons, only the first quadrant is depicted. In order to compare the penetration depth profiles with those from MDRANGE simulations (Fig. 6 in \cite {NordlundPRB}), our values were scaled with $1/\cos{\theta}$, since the mean ranges there do not refer to the direction of the surface normal, but correspond to the projection onto the direction of the incident H-ions.
    }    \label{fig:S1}
\end{figure}

The physical ($U_\text{AP}$, $\theta$) distributions of protons impinging on pad~2 and pad~3 of the SDD  are extracted from particle tracking simulations (see Fig.~S2). 
\begin{figure}
    \centering
    \renewcommand{\thefigure}{S2}
       \includegraphics[width=0.48\textwidth]{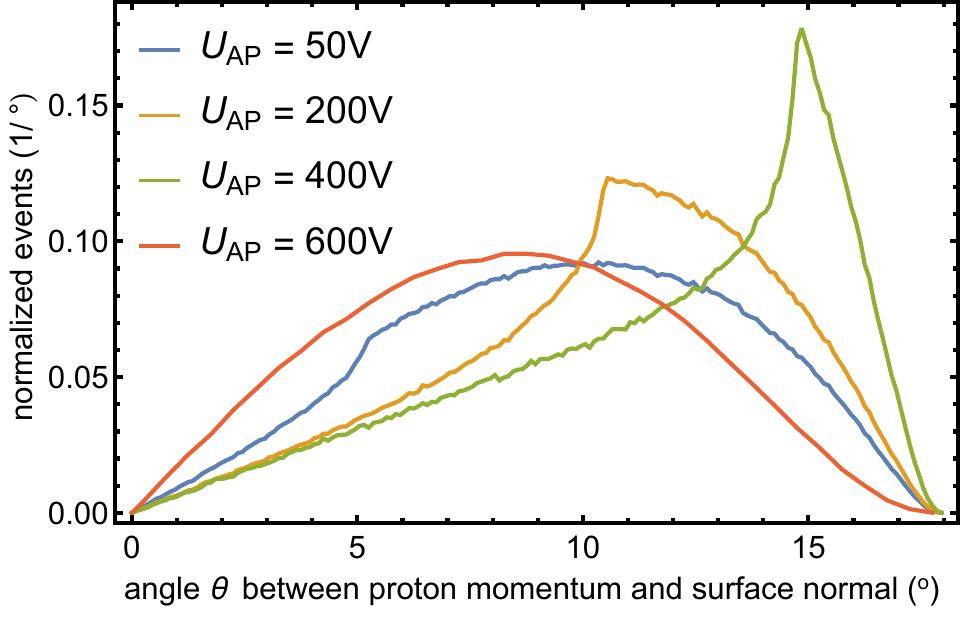}
    \caption{ Angular distributions of protons at given retardation voltages $U_\text{AP}$ impinging on pad~2 and pad~3 of the SDD. $\theta$ is the angle between the proton momentum and the surface normal. The total number of events for the individual $U_\text{AP}$ settings are normalized for better visibility.    
    }    \label{fig:S2}
\end{figure}
Protons with $\left<100\right>$ axial channeling incidence ($\theta \approx 0$) are rare. Moreover, they have to pass the amorphous overlayer of 30 nm  before hitting the active area of the SDD. As a consequence of beam spread due to multiple scattering \cite{Khalid}, protons experiences a reduced channeling (critical channeling angles in silicon for 15 keV protons : $\alpha_{\text{crit}}^{\left<100\right>}\leq 3^{o} $ \cite{Grahmann, Hobler}).  As the impact angle of protons spans the range of $0<\theta <20^\circ$ , channeling occurs preferentially on high-index axial and planar channels though the probability for the individual process is greatly reduced due to multiple scattering in the 30 nm thick $\text{Al}/\text{Al}_{2}\text{O}_{3}$ overlayer. The influence of the thickness of the passivation layer on proton channeling is decisive, as W. Khalid \cite{Khalid}  has shown for detectors with thinner cover layers, e.g. 1 nm of $\text{Al}_{2}\text{O}_{3}$ or 3 nm of $\text{SiO}_2$, when using incident protons in the same energy range.\\

In our simulation of fractional losses \cite{a11} we have also stated that backscattered protons may return to the detector after motion reversal due to the electrostatic potential of the analyzing plane electrode of the aSPECT spectrometer. Those protons hit the detector again with the energy and angle to the normal they had when leaving the top layer. Therefore, all possible hits of a proton due to backscattering were taken into account by adding the collected charge from all hits in the active region of the detector. In the meantime, we have realized that this is only partially correct. As J. Phillips \cite{Phillips} has shown, when low-energy hydrogen ions ($\text{H}^+$) pass through materials such as aluminum or more precisely: the 4 nm thin alumina top layer, electrons are captured or lost from the ions and particles of positive ($\text{H}^+$), neutral ($\text{H}^0$), and negative charge ($\text{H}^-$) arise. The two latter ones do not hit the detector a second time. The energy dependence of the percentage of positive components ($\text{H}^+$) is tabulated in \cite{Phillips} and can be approximated by a straight line ($\text{H}^+= 8\% + E_\text{kin}^{'}\cdot 1.674 \%/\text{keV}$) in the energy range $4~\text{keV} <E_\text{kin}^{'} < 16~\text{keV}$ of interest.\\

\begin{figure}[h!]
    \centering
\renewcommand{\thefigure}{S3}
    \includegraphics[width=0.48\textwidth]{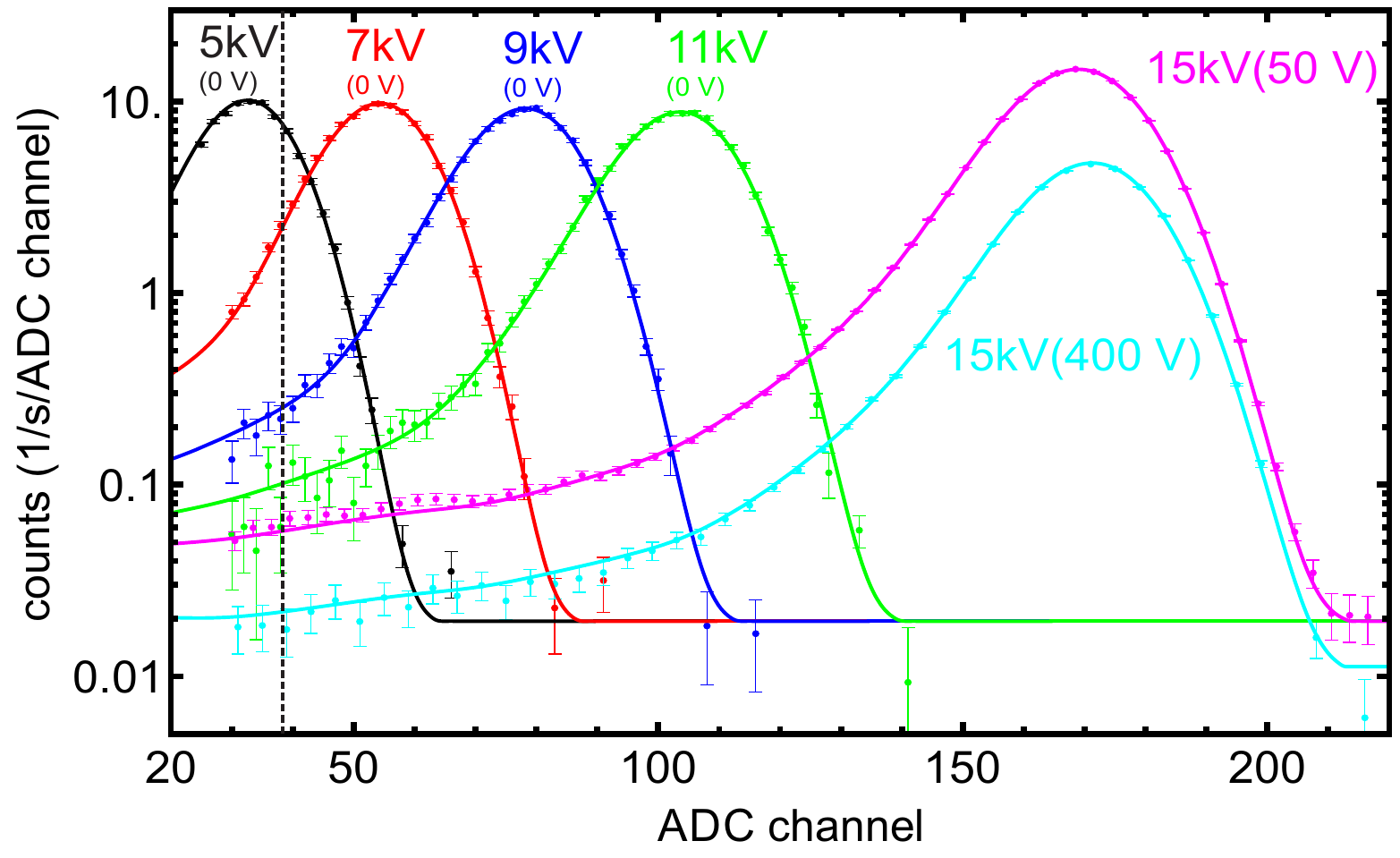}
    \caption{Measured pulse height spectra (pad 2) at different acceleration voltages $U_\text{acc}$ and retardation voltages $(U_\text{AP})$. The data were taken with a linear shaper (for more details see Ref.\cite{a11}). The corresponding histograms of the calculated pulse heights $\propto E_{\text{dep}}^{\text{C-TRIM}}$ are shown as continuous lines to improve
    readability. For pad 2, the proper threshold is indicated by the dashed line (ADC channel: 37.6). For the sake of completeness: the threshold ADC channel for pad 3 is 34.5 .  
    }    \label{fig:S3}
\end{figure}

In Fig.~S3, the experimental and simulated pulse height spectra are compared. The theoretical spectra were folded with a normalized Gaussian distribution (120~eV FWHM : 2.52~ADC channels). The FWHM-value corresponds to the intrinsic energy resolution derived in Ref. \cite{Lechner} for this type of SDD and includes the fluctuations of the charge collection efficiency parameters [see Eq.~(S.1)] from their mean values. From the parameter fit to the experimental pulse height spectra we obtain the two global parameters for all spectra, i.e., the conversion factor for converting ADC channels to deposit energy inside the detector and the scaling factor for the y-axis (detected counts). 
Furthermore, we determine from the fit the parameters of the  charge-collection
efficiency being S = 0.4616(21), L = 7.57(43) nm, c = 1.746(50), and $\tau$ = 37.35(37) nm. These values differ to a certain extent from the ones published in \cite{a11}, where the ionization depth profiles used were determined by the TRIM code. 

Finally, this procedure allows us to calculate the below-threshold losses (threshold ADC channel for pad 2: 37.6; dashed line in Fig.~S3) including the events with no energy deposition inside the detector. Figure~1 in the main text shows the fraction of undetected protons of the integral proton spectrum whose corresponding pulse heights fall below the threshold of the DAQ system. \\

\end{document}